\documentclass[twocolumn,superscriptaddress,showpacs,aps,pra]{revtex4-1}
\usepackage{graphicx}
\usepackage{dcolumn}
\usepackage{bm}
\usepackage[dvipdfm,colorlinks,linkcolor=blue, urlcolor=blue, anchorcolor=red, citecolor=blue]{hyperref}
\usepackage{appendix}
\usepackage{amsmath,amsfonts,amssymb}
\usepackage{booktabs}
\usepackage{float}
\usepackage{graphicx}
\usepackage{ulem}

\begin{document}
\title{One-way multiple beam splitter designed by quantum-like shortcut-to-adiabatic passage}

\author{Jiahui Zhang}\email[]{y10220159@mail.ecust.edu.cn}
\affiliation{School of Physics, East China University of Science and Technology, Shanghai 200237, China}
\author{Yating Wei}
\affiliation{School of Physics and Frontiers Science Center for Mobile Information Communication and Security, Southeast University, Nanjing, 211189, China}

\author{Li Deng}%
\email[]{dengli@ecust.edu.cn}
\affiliation{School of Physics, East China University of Science and Technology, Shanghai 200237, China}

\author{Yueping Niu}
\affiliation{School of Physics, East China University of Science and Technology, Shanghai 200237, China}
\affiliation{Shanghai Frontiers Science Center of Optogenetic Techniques for Cell Metabolism, East China University of Science and Technology, Shanghai 200237, China}
\affiliation{Shanghai Engineering Research Center of Hierarchical Nanomaterials, East China University of Science and Technology, Shanghai 200237, China}

\author{Shangqing Gong}
\email[]{sqgong@ecust.edu.cn}
\affiliation{School of Physics, East China University of Science and Technology, Shanghai 200237, China}
\affiliation{Shanghai Frontiers Science Center of Optogenetic Techniques for Cell Metabolism, East China University of Science and Technology, Shanghai 200237, China}
\affiliation{Shanghai Engineering Research Center of Hierarchical Nanomaterials, East China University of Science and Technology, Shanghai 200237, China}

\begin{abstract}
In this work, we introduce quantum-mechanical ``shortcut-to-adiabatic passage" (STAP) into the design of multiple beam splitter. This device consists of one input and $N$ output waveguide channels, which are connected via a mediator waveguide (WG). To implement ``STAP" but without additional couplings, this $(N+2)$-WG system is first reduced to a controllable $3$-WGs counterpart by ``Morris-Shore" transformation. Consequently, the reduced system can be directly compatible with all possible three-level ``STAP" methods. The results show that this novel design can achieve arbitrary ratio of multiple beam splitting and can significantly shorten the length of the device, which expands the application of ``STAP" in classical system and provides a direct visualization in space of typical ultrafast phenomena in time.  More uniquely, this new design exhibits a one-way energy transport behavior, which provides a very unique platform to demonstrate one-way localized ``STAP" process. The complete physical explanation of the underlying mechanism is presented. These excellent features may have profound impacts on exploring quantum technologies for promoting advanced optical devices.
\end{abstract}
\maketitle

\section{\label{sec:level1}Introduction}
Adiabatic passage (AP), initially proposed to achieve efficient and robust population transfer between quantum states, has been widely explored and applied in quantum optics and integrated optics~\cite{https://doi.org/10.1002/lpor.200810055, Menchon_Enrich_2016}. When carrying out classical analogs of the AP, the well-known STIRAP and its variants were most investigated due to their advantages of being robust, simple, and efficient~\cite{Shore:17}. Examples includes analogies of STIRAP in a three-WG directional coupler~\cite{PhysRevE.73.026607, Paspalakis}, 
fractional-STIRAP (F-STIRAP)~\cite{Dreisow2009} and multistate-STIRAP~
\cite{doi:10.1063/1.2828985}, to name a few. Among them, the design of beam splitter, which is used to split incident wave into two and multiple parts with the same or different intensities, is quite demanding, it has potential applications in optical circuits and communications~\cite{https://doi.org/10.1002/lpor.200810055}. While the $1\times2$ optical splitters are used routinely, e.g., Y-branch and T-branch junctions, the rapidly growing need for space-division multiplexing in optical transmission~\cite{Puttnam:21}, computing~\cite{10.3389/fphy.2020.589504} and sensing networks~\cite{Polino:19} translates into the need for multi-port splitters~\cite{Franz_2021}.

Generally adiabatic devices profit from similar properties as AP but suffer from a fundamental limitation in terms of speed, either in time (such as atomic system), or in space (such as photonic system). In order to meet the necessary adiabatic conditions, adiabatic devices require more space and resources, which will reduce device density and affect its performance. 
To address this problem, a range of methods referred as ``Shortcut-to-Adiabatic passage" (STAP) \cite{TORRONTEGUI2013117, RevModPhys.91.045001}, which was initially proposed to accelerate adiabatic quantum-state transfer~\cite{Demirplak2003, Berry_2009, PhysRevLett.105.123003, Li2021, Zhang_2021, Dou2021}, 
has recently been applied to design WG devices~\cite{Tseng:12, Lin:12, PhysRevA.90.023811, Martinez-Garaot:14, PhysRevA.91.053406, Pan:15, PhysRevApplied.18.014038, Evangelakos2023}.
It presents significant advantages with respect to conventional STIRAP-like schemes, such as relaxing the demands of the exact evolution. Integrated optics waveguide platform enabled by the advanced micro/nano fabrication technology provides an ideal platform to realize the device geometries designed by various ``STAP" protocol~\cite{Chung_2019}. A review focusing on the applications of ``STAP" methods in optical WGs can be found in Ref.~\cite{doi:10.1080/23746149.2021.1894978}.
Despite numerous achievements in the field of optics, the application of ``STAP" in multiple beam splitter has not been widely reported to date, especially its ability to obtain arbitrary beam splitting and achieve one-way energy transport.

In this paper, by performing an analog of the quantum-mechanical ``STAP" in classic optical system, a theoretical method for achieving efficient multiple beam splitting is proposed. This device consists of one input and $N$ output WG channels, which are connected via a mediator WG. With the modified inter-WG coupling strengths, the behaviors of beam splitting with arbitrary ratio as well as complete energy transfer are capable of realizing in the same coupler by selecting different input ports. The results prove that this coupled-WG system can reduce the length of device and exhibit a one-way energy transport behavior. Furthermore, if we consider $N =1$, this device can also render a ``F-STIRAP"-like beam splitting. These properties make it possible to design some interesting WG devices.

\section{Theoretical analysis}\label{sec2}
\begin{figure}[h]
\centering
\includegraphics[width=0.475\textwidth]{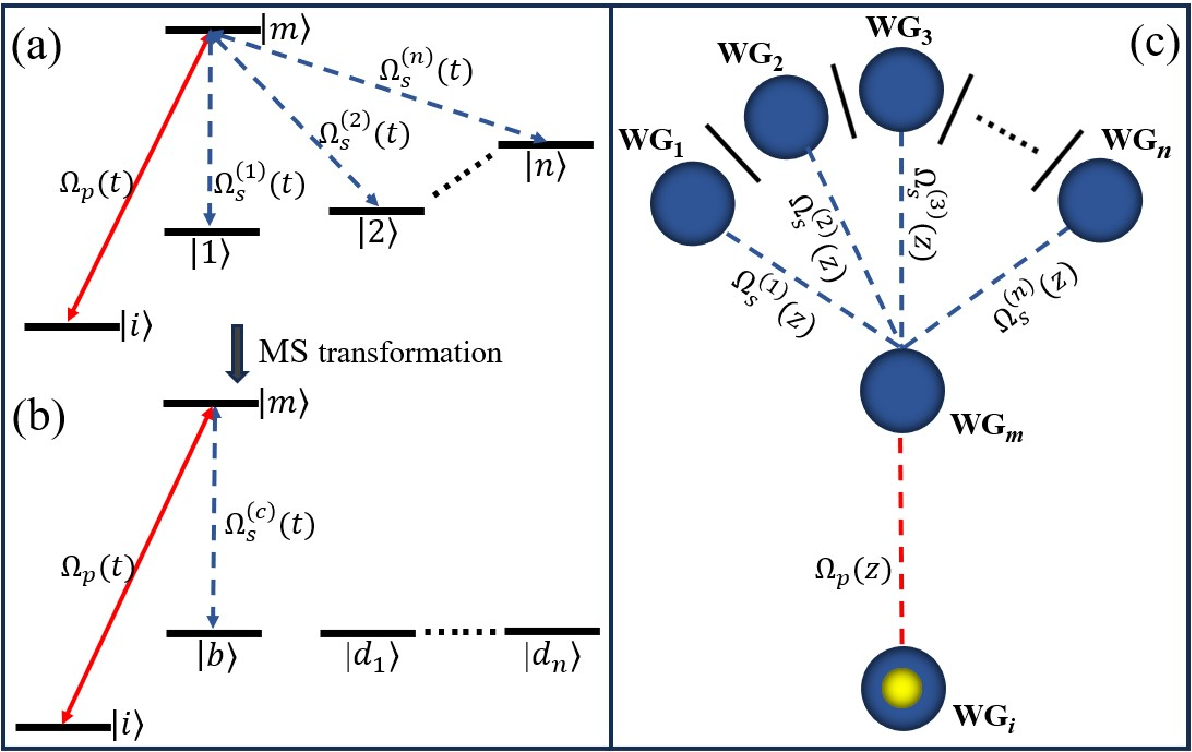}
\caption{(Color online) (a) Linkage pattern of the $N$-pod quantum system; (b) The effective three-level quantum counterpart after the MS transformation; (c) Cross section of a possible waveguide arrangement of the multiple beam splitter. According to the concept of ``STAP", this device contains a $(N+2)$ waveguide coupler with a novel structure, which has some specific curved structures for input/output WGs, and the middle WG is the straight structure. The dashed lines depict the couplings between the WGs, which are changed by changing the distance between the WGs~\cite{PhysRevA.85.055803, PhysRevA.103.053705}.}
\label{fig1}
\end{figure}

We start by considering the model of a $(N+2)$-level quantum system whose linkage pattern is depicted in Fig.~\ref{fig1}(a), in which the initial state $|i\rangle$ is coupled to the excited state $|m\rangle$ with a pump pulse $\Omega_P(t)$, and the other $N$ ground states ${|1\rangle, |2\rangle,...,|n\rangle}$ are also coupled to the excited state with a battery of Stokes pulses [$\Omega_S^{(1)}(t), \Omega_S^{(2)}(t)\cdot\cdot\cdot\Omega_S^{(n)}(t)$]. 
The evolution can be described by the time-dependent Schr\"{o}dinger equation:
\begin{eqnarray}\label{1}
i\hbar\displaystyle\frac{d}{d t} \psi(t)=H(t) \psi(t),
\end{eqnarray}
where $\psi(t)=[\psi_{i}(t), \psi_{m}(z), \psi_{1}(z),..., \psi_{n}(z)]^T$ with $\psi_{k}(t)$ being the probability
amplitude of the state $| k\rangle (k=i, m, 1,...,n)$.
Under the rotating-wave approximation, the Hamiltonian $H(t)$ is written as
\begin{eqnarray}\label{2}
H(t)=
\left(
\begin{array}{ccccc}
0&\Omega_{P}(t)&0&\cdots&0\\
\Omega_{P}(t)&0&\Omega_{S}^{(1)}(t)&\cdots&\Omega_{S}^{(n)}(t)\\
0&\Omega_{S}^{(1)}(t)&0&\cdots&0\\
\vdots&\vdots&\vdots&\ddots&\vdots\\
0&\Omega_{S}^{(n)}(t)&0&0&0
\end{array}
\right).
\end{eqnarray}
Based on the ``Morris-Shore" transformation~\cite{PhysRevA.74.053402}, one can simplify the analysis by reducing the dynamics of this $(N+2)$-level quantum system into controllable three-level one, as illustrated in Fig.~\ref{fig1}(b). The transformed Hamiltonian under MS basis \{$|i\rangle$, $|m\rangle$, $|b\rangle$\} can be read as follows:
\begin{eqnarray}\label{3}
H_{T}(t)=
\left(
\begin{array}{ccc}
0&\Omega_{P}(t)&0\\
\Omega_{P}(t)&0&\Omega_{S}^{(c)}(t)\\
0&\Omega_{S}^{(c)}(t)&0\\
\end{array}
\right),
\end{eqnarray}
where a bright superposition $|b\rangle$ is coupled to $|m\rangle$ with a coupling which is the root-mean-square (rms) of the initial couplings,
\begin{equation}\label{4}
\Omega_{S}^{(c)}(t)=\sqrt{\sum_{j=1}^{n}[\Omega_{S}^{(j)}(t)]^2}.
\end{equation}
The bright superposition has the vector form
\begin{equation}\label{5}
|b\rangle=\frac{[0, 0, \Omega_{S}^{(1)}(t), \Omega_{S}^{(2)}(t),..., \Omega_{S}^{(N)}(t)]^T}{\Omega_{S}^{(c)}(t)}.
\end{equation}
Diagonalizing Hamiltonian~(\ref{3}), one can obtain a dark state
\begin{eqnarray}\label{6}
| \Phi_{0}(t)\rangle=
\left(
\begin{array}{ccc}
\cos\alpha(t)\\
0\\
-\sin\alpha(t)\\
\end{array}
\right),
\end{eqnarray}
where the mixing angle $\tan\alpha(t)=\Omega_{S}^{(c)}(t)/\Omega_{P}(t)$.
In principle, if adiabatic conditions are satisfied, a perfect population transfer from $|i\rangle$ to superposition $|b\rangle$ can be achieved via a STIRAP-like process~\cite{RevModPhys.89.015006}. However, a perfect quantum-state transfer needs to satisfy the adiabatic criterion. This means it requires a very long evolution time, analogous to requiring a long spatial variation of device. Therefore, it is necessary to speed-up the adiabatic procedure towards the perfect final outcome. Although a solution for implementing ``STAP" in $N$-pod system has recently been proposed~\cite{saadati2020quantum}, the proposal requires additional couplings among initial state $|i\rangle$ and the final ground states ${|1\rangle, |2\rangle,...,|n\rangle}$, which is generally difficult or even impossible for practical implementations. Particularly, the existence of imaginary coupling coefficients makes the proposal unrealistic.

To implement ``STAP" but without additional couplings, we can introduce a pair of counter-adiabatic pulses into $\Omega_{P}(t)$ and $\Omega_{S}^{(c)}(t)$, to find a pair of modified coupling pulses $\tilde{\Omega}_{P}(t)$ and $\tilde{\Omega}_{S}^{(c)}(t)$. In principle, such a pair of modified pulses can be obtained by several three-level ``STAP" methods, e.g., ``invariant-based inverse engineering"~\cite{PhysRevA.86.033405, 10.1063/5.0183063}, ``multiple Schr\"{o}dinger dynamics"~\cite{PhysRevLett.109.100403}, ``chosen paths"~\cite{Wu:17}, and ``superadiabatic dressed states"~\cite{Zhou2017}, which show different mathematical processes but similar underlying physics; see recent reviews~\cite{RevModPhys.91.045001, doi:10.1080/23746149.2021.1894978}.
The form of the modified Hamiltonian can be written as
\begin{eqnarray}\label{7}
\tilde{H}_{T}(t)=
\left(
\begin{array}{ccc}
0&\tilde{\Omega}_{P}(t)&0\\
\tilde{\Omega}_{P}(t)&0&\tilde{\Omega}_{S}^{(c)}(t)\\
0&\tilde{\Omega}_{S}^{(c)}(t)&0\\
\end{array}
\right).
\end{eqnarray}
Among the methods mentioned above, let us use the ``chosen paths" method as an example to demonstrate our treatment. This method supports the following dressed states
\begin{eqnarray}\label{8}
| \Phi_{0}(t)\rangle=
\left(
\begin{array}{ccc}
\cos\mu(t)\cos\theta(t)\\
i\sin\mu(t)\\
\cos\mu(t)\sin\theta(t)\\
\end{array}
\right),
\end{eqnarray}
\begin{eqnarray}\label{9}
| \Phi_{\pm}(t)\rangle=
\left(
\begin{array}{ccc}
\sin\theta(t)\mp i\sin\mu(t)\cos\theta(t)\\
\mp\cos\mu(t)\\
-\cos\theta(t)\mp i\sin\mu(t)\sin\theta(t)\\
\end{array}
\right)/\sqrt{2}.
\end{eqnarray}
It is convenient to move $\tilde{H}_{T}(t)$ to the frame with the time-independent chosen paths being basis by the unitary operator $U_0=\sum_{\xi=\pm,0}| \Phi'_{\xi}(t)\rangle\langle\Phi'_{\xi}(t)|$. Hamiltonian~(\ref{7}) in new frame becomes
\begin{eqnarray}
\tilde{H}'_{T}(t)&&=U_0(t)\tilde{H}_{T}(t)U^{\dag}_0(t)-iU_0(t)\dot{U}^{\dag}_0(t)\nonumber\\
&&=\Lambda(t)(| \Phi'_{+}(t)\rangle\langle\Phi'_{+}(t)|-| \Phi'_{-}(t)\rangle\langle\Phi'_{-}(t)|)\nonumber\\
&&+[\eta_{\pm}(t)| \Phi'_{\pm}(t)\rangle\langle\Phi'_{0}(t)|+H.c.].
\end{eqnarray}
It is not difficult to find that the interactions of state $| \Phi'_{0}(t)\rangle$ with the other two states can be eliminated only when $\eta_{\pm}(t)=0$.
Accordingly, one can easily derive that
\begin{eqnarray}
\tilde{\Omega}_{P}(t)&=-\dot{\theta}(t)\sin\theta(t)\cot\mu(t)-\dot{\mu}(t)\cos\theta(t),\nonumber \\
\tilde{\Omega}_{S}^{(c)}(t)&=\dot{\theta}(t)\cos\theta(t)\cot\mu(t)-\dot{\mu}(t)\sin\theta(t) \label{12}.
\end{eqnarray}

The state function of the system can be further mapped into the dressed-state space by~\cite{PhysRevLett.122.094501}
\begin{equation}\label{12}
| \Phi'(t)\rangle=\sum_{\xi=0, \pm}\tilde{A}_\xi(t)| \Phi'_{\xi}(t)\rangle,
\end{equation}
where $\tilde{A}_\xi(t)=a_\xi(t_0)\exp[-i\int_{t_0}^t\varepsilon_\xi(t')dt']$ with $a_\xi(t_0)$ being determined by the initial state $| \Phi'(t_0)\rangle$ and $\varepsilon_\xi(t)=\langle\Phi'_{\xi}(t)|H_{M}(t)|\Phi'_{\xi}(t)\rangle$.
If $a_0 (t_0)=1$ and $a_+(t_0)=a_-(t_0)=0$, the state function will follow the evolution of $| \Phi'_{0}(t)\rangle$. In this case, a perfect quantum state transfer from $|i\rangle$ to $|b\rangle$ can be achieved if we set $\mu(t)=\mu(t_f)=0$, and $\theta(t_0)= 0, \theta(t_f)=\pi/2$.
On the other hand, if $a_0 (t_0)=0$ and $a_+(t_0)=a_-(t_0)=1/\sqrt{2}$, the state function will follow the evolution of $| \Phi'_{\pm}(t)\rangle$.
Eq.~(\ref{12}) shows that the state transfer from $|b\rangle$ to $|m\rangle$ can be realized under the same conditions adopted in the state transfer from $|i\rangle$ to $|b\rangle$.
It is quite interesting that population transfer is realized from $|i\rangle$ to $|b\rangle$, while from $|b\rangle$ to $|m\rangle$ under the same condition, showing an asymmetric transfer behavior in this system.

Now let us go back to the $(N+2)$-level quantum system and design the modified pulses by
comparing the Hamiltonian~(\ref{3}) and~(\ref{7}). Like Eq.~(\ref{4}),
we impose~\cite{10.1063/5.0223526}
\begin{equation}\label{13}
\tilde{\Omega}_{S}^{(c)}(t)=\sqrt{\sum_{j=1}^{n}[\tilde{\Omega}_{S}^{(j)}(t)]^2},
\end{equation}
and calculate inversely the modified pulses as $\sum_{j=1}^{n}[\tilde{\Omega}_{S}^{(j)}(t)]^2=(\tilde{\Omega}_{S}^{(c)})^2(t)$,
which leads to a modified Hamiltonian with the same form as Hamiltonian~(\ref{2})
\begin{eqnarray}\label{14}
\tilde{H}(t)=
\left(
\begin{array}{ccccc}
0&\tilde{\Omega}_{P}(t)&0&\cdots&0\\
\tilde{\Omega}_{P}(t)&0&\tilde{\Omega}_{S}^{(1)}(t)&\cdots&\tilde{\Omega}_{S}^{(n)}(t)\\
0&\tilde{\Omega}_{S}^{(1)}(t)&0&\cdots&0\\
\vdots&\vdots&\vdots&\ddots&\vdots\\
0&\tilde{\Omega}_{S}^{(n)}(t)&0&0&0
\end{array}
\right).
\end{eqnarray}
Remarkably, the Hamiltonian~(\ref{14}) can describe other physical systems apart from the quantum system shown in Fig.~\ref{fig1}(a). For example it represents in the paraxial approximation and substituting time by a longitudinal coordinate $(N+2)$ coupled WGs~\cite{Huang:94}, where the couplings be controlled by waveguide separation. In particular $\tilde{\Omega}_{P}(t\rightarrow z)$ and $\tilde{\Omega}_{S}^{(n)} (t\rightarrow z)$ may be manipulated to split the input wave into two output channels~\cite{Tang_2023, SHI201629} or multiple output channels~\cite{PhysRevA.85.055803}.

By mapping the above results into spatial dimension~\cite{Huang:94}, as sketched in Fig.~\ref{fig1}(c), our analogy in an optical system is proposed by a spatial $(N+2)$-WG coupler. Fig.~\ref{fig1}(c) shows a cross section of the possible geometry of this device, which consists of one input WG$_i$, one mediator WG$_m$, and finite number of $N$ output WG$_n (n=1,2,...,N)$. Here, WG$_i$, WG$_m$ and WG$_n (n=1,2,...,N)$ correspond to the quantum states $|i\rangle$, $|m\rangle$ and $|n\rangle (n=1,2,...,N)$, respectively, and WG$_m$ serves as an intermediate WG spatially coupled with WG$_i$ and WG$_n (n=1,2,...,N)$. According to the concept of ``STAP", our device has a novel structure, the input/output WGs have some specific curved structure, and the middle WG is the straight structure. The dashed lines in the diagram depict the couplings between the WGs, which depend strongly on the distance between the WGs due to the evanescent-wave nature of the waveguide couplings. The WGs of the output channels should not be coupled to each other but only to the mediator WG$_m$; hence they are supposed to be isolated from each other~\cite{PhysRevA.85.055803, Tang_2023}. Obviously, this device can only be effectively simulated using WGs in three dimensional arrangements, which can be fabricated by the femtosecond laser direct-write technique~\cite{Szameit_2010}.

The evolution of the wave amplitudes can be described by a set of $(N+2)$ coupled differential equations~\cite{Huang:94},
\begin{equation}\label{15}
i\frac{d}{dz}A(z)=M(z)A(z),
\end{equation}
where $A=[a_{i}(z),a_{m}(z),a_{1}(z),..., a_{n}(z)]^T$, $a_{k}(z) (k=i, m, 1,..., n)$ is the light amplitude in the $k$th waveguide and the dimensionless light intensity is $I_k= |a_k(z)|^2$. The coupling matrix shown below:
\begin{eqnarray}\label{16}
M(z)=
\left(
\begin{array}{ccccc}
0&\tilde{\Omega}_{P}(z)&0&\cdots&0\\
\tilde{\Omega}_{P}(z)&0&\tilde{\Omega}_{S}^{(1)}(z)&\cdots&\tilde{\Omega}_{S}^{(n)}(z)\\
0&\tilde{\Omega}_{S}^{(1)}(z)&0&\cdots&0\\
\vdots&\vdots&\vdots&\ddots&\vdots\\
0&\tilde{\Omega}_{S}^{(n)}(z)&0&0&0
\end{array}
\right),
\end{eqnarray}
the coupling strengths can be expressed as
\begin{eqnarray}
\tilde{\Omega}_{P}(z)&&=-\dot{\theta}(z)\sin\theta(z)\cot\mu(z)-\dot{\mu}(z)\cos\theta(z),\nonumber \\
\tilde{\Omega}_{S}^{(c)}(z)&&=\dot{\theta}(z)\cos\theta(z)\cot\mu(z)-\dot{\mu}(z)\sin\theta(z),\label{17}
\end{eqnarray}
in which $\sum_{j=1}^{n}[\tilde{\Omega}_{S}^{(j)}(z)]^2=(\tilde{\Omega}_{S}^{(c)})^2(z)$. In typical implementations where the coupling strengths can be modulated by changing the distance between the WGs.

Based on the above analysis, the boundary conditions $\mu(z)$ and $\theta(z)$ should satisfy~\cite{PhysRevA.109.023109}
\begin{equation}
\mu(z_0)=\mu(z_f)=0, \quad
\theta(z_0)=0, \quad \theta(z_f)=\pi/2\label{18}.
\end{equation}
To meet these conditions,
$\mu(z)$ and $\theta(z)$ can be set as
\begin{eqnarray}
\theta(z)&&=\frac{\pi z}{2z_f}-\frac{1}{3}\sin(\frac{2\pi z}{z_f})+\frac{1}{24}\sin(\frac{4\pi z}{z_f}),\nonumber \\
\mu(z)&&=\frac{\zeta}{2}\left[1-\cos(\frac{2\pi z}{z_f})\right]\label{19}.
\end{eqnarray}
In principle, if light wave is incident of WG$_{i}$, similar to Eq.~(\ref{8}), it follows:
\begin{eqnarray}\label{20}
| \Phi'_{0}(z)\rangle=
\left(
\begin{array}{ccc}
\cos\mu(z)\cos\theta(z)\\
i\sin\mu(z)\\
\cos\mu(z)\sin\theta(z)\\
\end{array}
\right),
\end{eqnarray}
the output port is expected to be WG$_{n}$, the power distribution of each output port can be customized arbitrarily by utilizing respective space-varying coupling $\tilde{\Omega}_{S}^{(n)}(z)$;
if light waves are incident from WG$_{n} (n=1, 2,\ldots, N)$ simultaneously, similar to Eq.~(\ref{9}), it follows:
\begin{eqnarray}\label{21}
| \Phi'_{0}(z)\rangle=
\left(
\begin{array}{ccc}
\sin\theta(z)\mp i\sin\mu(z)\cos\theta(z)\\
\mp\cos\mu(z)\\
-\cos\theta(z)\mp i\sin\mu(z)\sin\theta(z)\\
\end{array}
\right)/\sqrt{2},
\end{eqnarray}
the output port is expected to be WG$_{m}$.
\begin{figure*}
\centering{\includegraphics[width=0.9\textwidth]{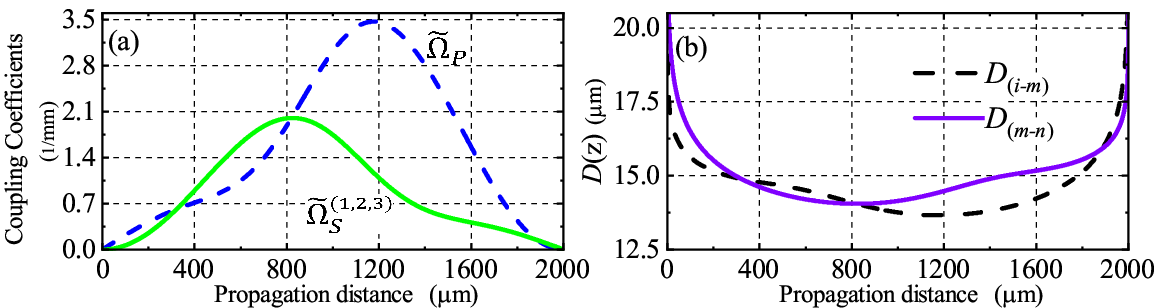}}
\caption{(Color online) (a) Coupling coefficients as a function of $z$. (b) The distance functions of input/middle and middle/output WGs. According to the relationship between coupling strength and distance, we can transfer the coupling strength to the geometrical structure. Note that in the simulation, the device length is chosen to be $2000\mu m$ and the default widths for the WG$_i$, WG$_m$ and WG$_n$ are chosen to be $2\mu m$, and the other parameters are chosen as $\zeta=0.145\pi, \Omega_0=4.387 mm^{-1}, \gamma\simeq1.409\mu m^{-1}, D_0 =3.687\mu m$~\cite{Chen_2018}.}
 \label{fig2}
\end{figure*}
Here, let us take the equal proportional beam splitting of WG$_{i}$$\rightarrow$WG$_{n} (n=1, 2, 3)$ as example to describe the geometry of the device.
The modified couplings $\tilde{\Omega}_P(z)$ and $\tilde{\Omega}_{S}^{(1, 2, 3)}(z)$ are illustrated in Fig.~\ref{fig2}(a).
As we can see, the coupling strengths $\tilde{\Omega}_P(z)$ and $\tilde{\Omega}_{S}^{(1, 2, 3)}(z)$ are alternately dominant along the propagation direction, and the latter couplings have the same values. Thus WG$_1$, WG$_2$ and WG$_3$ are arranged in a circular geometry, with WG$_m$ at the center of the circle, as shown in Fig.~\ref{fig1}(c).
According to coupled mode theory (CMT) \cite{Huang:94}, the relation between
the coupling strengths and waveguide separation can be well fitted by the exponential relations~\cite{Chen_2018}:
\begin{eqnarray}
\tilde{\Omega}_P(z)&&=-\Omega_0\exp[-\gamma(D_{(i-m)}(z)-D_0)],\nonumber \\
\tilde{\Omega}_{S}^{(1, 2, 3)}(z)&&=-\Omega_0\exp[-\gamma(D_{(m-n)}(z)-D_0)],\label{22}
\end{eqnarray}
in which  $D_{(i-m)}$ and $D_{(m-n)}$ stand for the separations between WG$_{i}$ and WG$_{m}$ and between WG$_{m}$ and WG$_{n}$, respectively. Therefore, we can design the function of separation between WGs to schematize the function of coupling strengths $\Omega_P(z)$ and $\tilde{\Omega}_{S}^{(c)}(z)$.
The corresponding device parameters are shown in Fig.~\ref{fig2}(b).

\section{Result and Discussion}
Below, let us still limit the model to $N=3$ for an explicit demonstration. For simplicity, the couplings among WG$_m$ and WG$_n (n=1, 2, 3)$ are assumed to share the same $z$-dependence but they may have in general different magnitudes, i.e., $\tilde{\Omega}_{S}^{(1)}(z):\tilde{\Omega}_{S}^{(2)}(z):\tilde{\Omega}_{S}^{(3)}(z)=\eta_1:\eta_2:\eta_3$. 
The post-coupling power distribution along each WG can be obtained by solving the coupled mode equation~(\ref{15}) with the coupling matrix~(\ref{16}).
\begin{figure*}[t]
\centering{\includegraphics[width=1\textwidth]{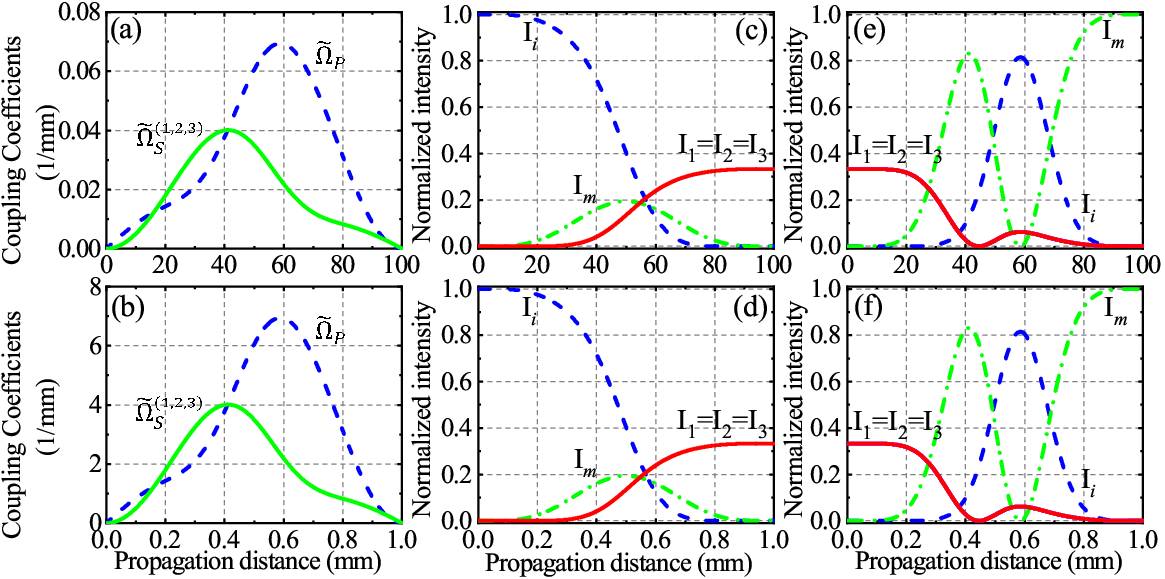}}
\caption{The equal multiple beam splitter designed by ``STAP". Left column: coupling strengths as a function of $z$;
Middle column: $1: 1: 1$ beam splitting for $\eta_1=\eta_2=\eta_3$; Right column: one-way optical energy transfer. Top frame: $z\in[0, 100]mm$; Bottom frame: $z\in[0, 1]mm$. Parameter used: $\zeta=0.145\pi$.}
 \label{fig3}
\end{figure*}

In Fig.~\ref{fig3} (left column), we show the well-tailored inter-WG coupling strengths as a function of $z$, which are used for equal multiple beam splitting.
The blue dash, green dot and red solid lines in Fig.~\ref{fig3}(c) illustrate the simulated light intensity distributions along WG$_i$, WG$_m$, WG$_n (n=1, 2, 3)$, respectively, it can be observed that the light wave input from WG$_i$ can be equally split to WG$_1$, WG$_2$ and WG$_3$ with $z=100mm$.
As it is observed, the mediator WG$_m$ carries a little transient light during the propagation, which will undoubtedly reduce transmission efficiency. Fortunately,
due to the unique property of the ``STAP", one can reproduce the same perfect efficiencies of multiple beam splitter via a shorter length, as illustrated in Fig.~\ref{fig3}(d). Clearly, the device length is reduced to $1mm$ in this numerical example, corresponding to a significant reduction in device length as compared to $100mm$. This is essential for the miniaturization of the device and the reduction of losses during transmission.

Another interesting feature is that the multiple beam splitter designed by ``STAP" can present a one-way energy transport behavior, that is, if light is incident WG$_n (n=1, 2, 3)$ simultaneously, a complete energy transfer to WG$_m$ instead of getting transferred back to WG$_i$ can be achieved, as shown in Figs.~\ref{fig3}(e) and~\ref{fig3}(f), which provides additional control over the distribution of light in the WGs. The above asymmetric behavior of energy transfer can be explained by two different evolution processes~\cite{Tang_2023}. For the left port of WG$_i$ incidence, the evolution follows Eq.~(\ref{20}), thus leading to a result of $[I_i(z_0), 0, 0, 0, 0]\rightarrow[0, 0, \frac{1}{3}I{_1}(z_f), \frac{1}{3}I{_2}(z_f), \frac{1}{3}I{_3}(z_f)]$ under the conditions of~(\ref{18}). For the left port of WG$_n (n=1, 2, 3)$ simultaneous incidence, the evolution follows Eq.~(\ref{21}), exhibiting another result of $[0, 0, \frac{1}{3}I{_1}(z_0), \frac{1}{3}I{_2}(z_0), \frac{1}{3}I{_3}(z_0)]\rightarrow[0, I{_m}(z_f), 0, 0, 0]$ under the same boundary conditions.
Indeed, the output port for incidence from WG$_n$ is closely related to the value of $\zeta$, while the output port for incidence from WG$_i$ is immune to the value of $\zeta$. The detailed explanations are presented in Appendix~\ref{app1}. Additionally, if we assume that $N=1$, which represents a system of three evanescently coupled WGs, see Appendix~\ref{app2}.

\begin{figure*}[t]
\centering{\includegraphics[width=1\textwidth]{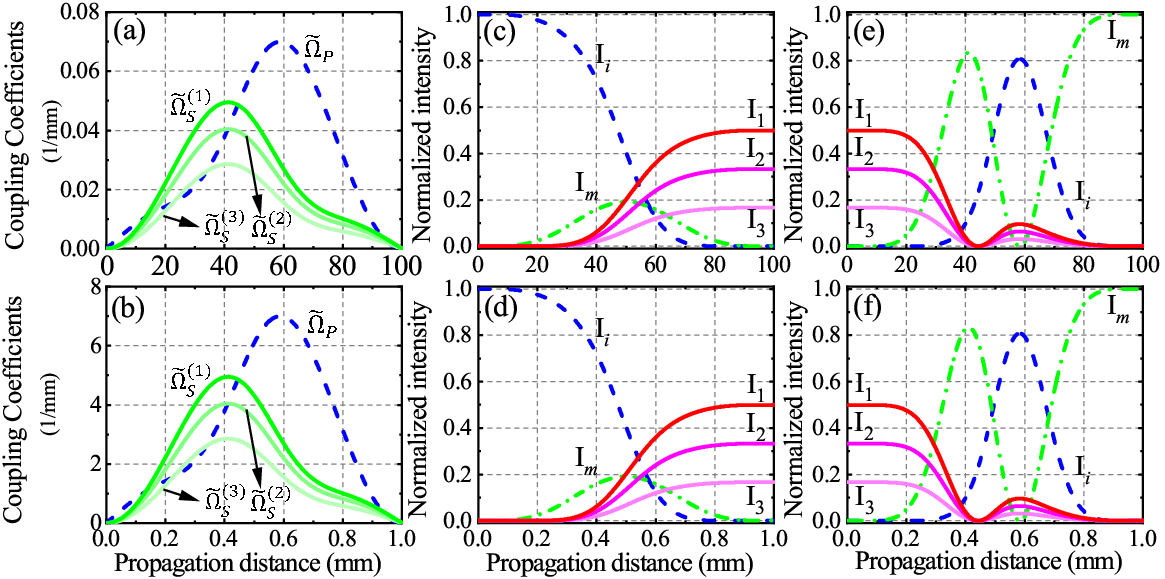}}
\caption{The variable multiple beam splitter designed by ``STAP". Left column: coupling strengths as a function of $z$; Middle column: $3: 2: 1$ beam splitting for $\eta_1: \eta_2: \eta_3=\sqrt{3}: \sqrt{2}: 1$; Right column: one-way optical energy transfer. Top frame: $z\in[0, 100]mm$; Bottom frame: $z\in[0, 1]mm$. Parameter used: $\zeta=0.145\pi$.}
 \label{fig4}
\end{figure*}
More interestingly, this device can achieve arbitrary multiple beam splitting for WG$_i$ incidence, while no apparent intensity from WG$_i$ can be obtained for WG$_n (n=1, 2, 3)$ simultaneous incidence, as shown in Fig.~\ref{fig4}.
The corresponding modified inter-WG coupling strengths as a function of $z$ are shown in the left column of Fig.~\ref{fig4}. Figures.~\ref{fig4}(c) and~\ref{fig4}(d) illustrate the simulated light intensity distributions with $z=100mm$ and $1mm$, respectively. It can be observed that the light wave input from WG$_i$ can be completely transfer to WG$_1$, WG$_2$ and WG$_3$ in a ratio of $3:2:1$ for $\eta_1: \eta_2: \eta_3=\sqrt{3}: \sqrt{2}: 1$, which implies that the results of $[I{_i}(z_0), 0, 0, 0, 0]\rightarrow[0, 0, \frac{3}{6}I{_1}(z_f), \frac{2}{6}I{_2}(z_f), \frac{1}{6}I{_3}(z_f)]$ has been realized. If light is incident from WG$_{n} (n=1, 2, 3)$ simultaneously, a complete energy transfer to WG$_m$ can be realized, this interesting phenomenon of $[0, 0, \frac{3}{6}I{_1}(z_0), \frac{2}{6}I{_2}(z_0), \frac{1}{6}I{_3}(z_0)]\rightarrow[0, I{_m}(z_f), 0, 0, 0]$ as shown in Figs.~\ref{fig4}(e) and~\ref{fig4}(f).


\section{Conclusion}\label{sec13}
In conclusion, a theoretical method is exhibited in this paper for multiple beam splitting by a quantum-like ``STAP". By coupling an input and $N$ output WGs with a mediator WG in space, efficient and robust multiple beam splitting can be achieved and the ratio of intensity can be customized arbitrarily by altering the space-dependent coupling strengths. This approach is versatile, in the sense that it can be compatible with several three-level ``STAP" methods. This novel design can significantly reduce the length of the device and can present a one-way energy transport behavior, these excellent features may have profound impacts on exploring quantum technologies for promoting optical devices with simple configuration and excellent performance. In addition, due to the analogies between quantum mechanics and wave optics, our protocol could provide a unique platform for optical simulation of the evolution of the particle wave function in a tripod system or a multi-pod system in quantum optics.

\appendix
\section{One-way wave propagation feature}
\label{app1}
\begin{figure}
\centering{\includegraphics[width=0.475\textwidth]{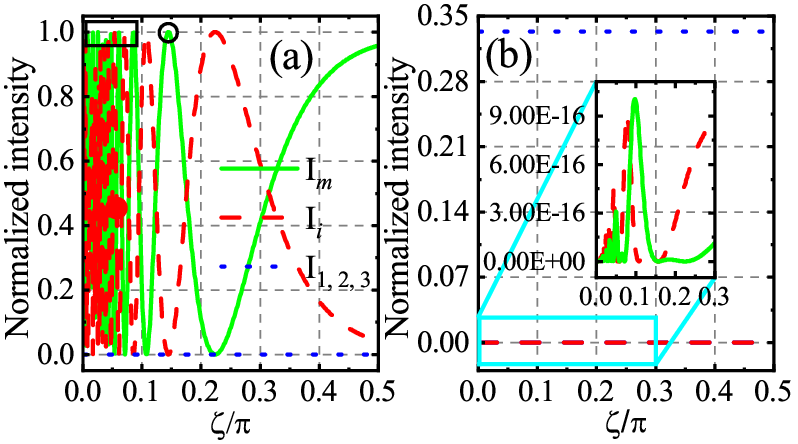}}
\caption{(a) The post-coupling intensity distributions as a function of $\zeta$ for WG$_n (n=1, 2, 3)$ simultaneous incidence; (b) The post-coupling intensity distributions as a function of $\zeta$ for WG$_i$ incidence. Note the parameter $\zeta\in(0.001, 0.499)\pi$, the step size is set to $0.001\pi$. All the other parameters are the same as those in~Fig.~\ref{fig3}(a).
}
 \label{fig5}
\end{figure}
In this appendix, we will show how to realize the one-way wave propagation feature via reasonably choosing the value of $\zeta$. Notably, $\zeta$ should $\in(0, \frac{\pi}{2})$ as an exact zero/$\frac{\pi}{2}$ value implies infinite/infinitesimal coupling strengths according to Eqs.~(\ref{17})-(\ref{19}). Figure.~\ref{fig5}(a) shows the post-coupling intensity distributions as a function of $\zeta$ for WG$_n (n=1, 2, 3)$ simultaneous incidence. The result clearly shows the output intensity of the propagation wave along WG$_i$ and  WG$_m$ will funnel back and forth as the value of $\zeta$ changes while that along WG$_n (n=1, 2, 3)$ can be ignored. The black circle corresponds to $\zeta=0.145\pi$, depicted in Fig.~\ref{fig3}(e), which illustrates that a one-way wave propagation feature. Furthermore, it is easy to find additional viable values, as shown in the area in the black box, which offer a variety of options for the design of one-way multiple beam splitter. In contrast, the output port for incidence from WG$_i$ is independent of the parameter $\zeta$, as illustrated in Fig.~\ref{fig5}(b). In this sense, this proposal provides a very unique platform to demonstrate one-way localized ``STAP" process.
\section{Three-waveguide coupler}\label{app2}
\begin{figure*}
\centering{\includegraphics[width=1\textwidth]{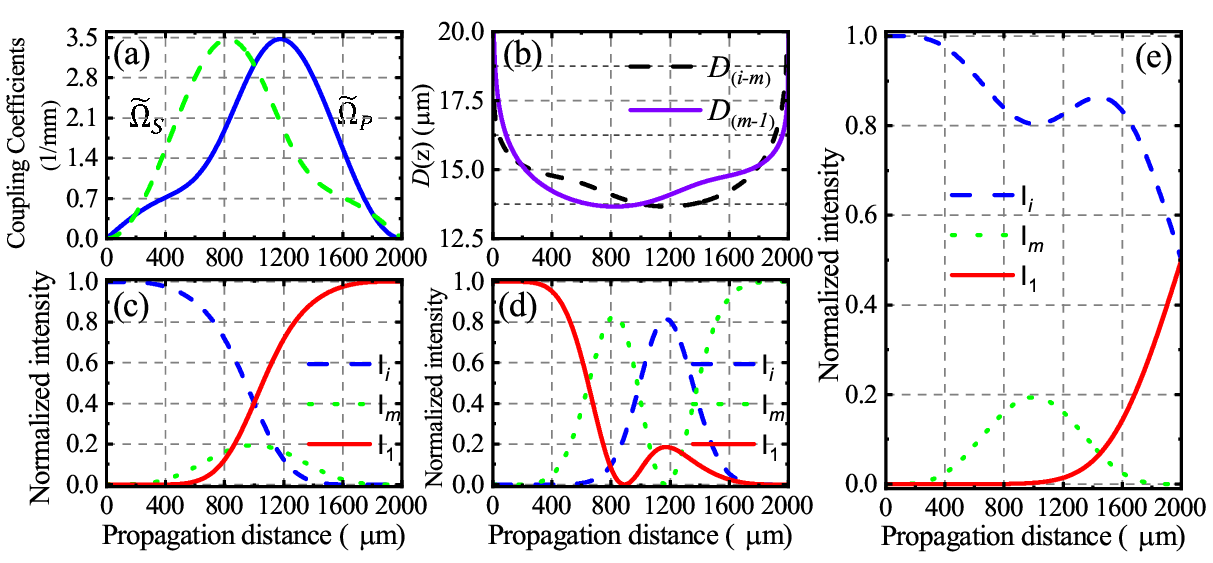}}
\caption{(a) The modified coupling coefficients as a function of $z$. (b) The distance functions of WG$_i$/WG$_m$ and WG$_m$/WG$_1$. All parameters are the same as those in Fig.~\ref{fig2}. (c) and (d) show the intensity distributions along the coupler designed by ``STAP". (e) shows an example of $1:1$ beam splitting, the $\theta(z)$ and $\mu(z)$ should be altered to be $\theta(z)=\pi z/4z_f-1/3\sin(\pi z/z_f)+1/24\sin(2\pi z/z_f), \mu(z)=\zeta/2\left[1-\cos(2\pi z/z_f)\right]$.}
 \label{fig6}
\end{figure*}
In this appendix, we consider the case of $N=1$, where the Hamiltonian~(\ref{7}) represents in the paraxial approximation and substituting time by a longitudinal coordinate three evanescently coupled WGs~\cite{Menchon-Enrich2013}, where the two outer WGs, WG$_i$ and WG$_1$ can both be used as input ports and the central one is WG$_m$. By performing quantum-classical analogs, $\tilde{\Omega}_{P}(z)$ [also $\tilde{\Omega}_{S}(z)$] in Hamiltonian~(\ref{7}) stands for the coupling strength between WG$_i$ and WG$_m$ (also WG$_m$ and WG$_1$). The propagation of light in this device can be modeled using standard CMT:
\begin{eqnarray}
i\frac{d}{dz}
\left(
\begin{array}{ccc}
a_i\\
a_m\\
a_1\\
\end{array}
\right)
=
\left(
\begin{array}{ccc}
0&\tilde{\Omega}_{P}(z)&0\\
\tilde{\Omega}_{P}(z)&0&\tilde{\Omega}_{S}(z)\\
0&\tilde{\Omega}_{S}(z)&0\\
\end{array}
\right)
\left(
\begin{array}{ccc}
a_i\\
a_m\\
a_1\\
\end{array}
\right).
\end{eqnarray}
Based on the exponential relationship, we can design the function of distance between WGs to schematize the function of coupling strength. 
The modified couplings $\tilde{\Omega}_{P}(z)$ and $\tilde{\Omega}_{S}(z)$ are illustrated in Fig.~\ref{fig6}(a). As we can see, the coupling strength is relatively tedious compared to the traditional Gaussian form, which requires a complicated curve of the input/output WG. To overcome this engineering problem, one
can employ ion implantation technology to engineer the arbitrary complicated curves of input/output WGs~\cite{PhysRevB.76.201101}. Figure.~\ref{fig6}(b) demonstrates the distance functions of WG$_i$/WG$_m$ and WG$_m$/WG$_1$. Based on this geometrical structure, we can easily fabricate our device with the ion implementation technique. The evolution of light energy transfer from the WG$_i$ to WG$_1$ (also from the WG$_1$ to WG$_i$) are shown in Figs.~\ref{fig6}(c) and~\ref{fig6}(d). The results indicate that this coupled-WG system exhibits a one-way energy transport behavior. 
In addition, if we reset the boundary conditions to
\begin{equation}
\mu(z_0)=\mu(z_f)=0,
\quad \theta(z_0)=0, \quad  \theta(z_f)=\pi/4,
\end{equation}
this three-WG coupler can render a ``F-STIRAP"-like beam splitting~\cite{Dreisow2009}, as shown in Fig.~\ref{fig6}(e). As we can see, a $50\%$ beam splitting between WG$_i$ and WG$_1$ can be achieved. Of course, with reasonable modification of the boundary conditions, any proportion of beam splitting can also be achieved.

\bibliography{reference}
\end{document}